# Cluster Accretion Shocks as Possible Acceleration Sites for Ultra High Energy Protons below the Greisen Cutoff

Hyesung Kang

Department of Earth Sciences, Pusan National University, Pusan 609-735, Korea;
e-mail: kang@astrophys.es.pusan.ac.kr

Dongsu Ryu

Dept. of Astronomy & Space Sci., Chungnam National Univ., Daejeon 305-764, Korea;
e-mail: ryu@sirius.chungnam.ac.kr

and

T. W. Jones

Department of Astronomy, University of Minnesota, Minneapolis, MN 55455;
e-mail: twj@astro.spa.umn.edu

## ABSTRACT

Three-dimensional hydrodynamic simulations of large scale structure in the Universe have shown that accretion shocks form during the gravitational collapse of one-dimensional caustics, and that clusters of galaxies formed at intersections of the caustics are surrounded by these accretion shocks. Estimated speed and curvature radius of the shocks are 1000-3000 km s$^{-1}$ and about 5 Mpc, respectively, in the $\Omega = 1$ CDM universe. Assuming that energetic protons are accelerated by these accretion shocks via the first-order Fermi process and modeling particle transport around the shocks through Bohm diffusion, we suggest that protons can be accelerated up to the *Greisen cutoff energy* near $6 \times 10^{19}$ eV, provided the mean magnetic field strength in the region around the shocks is at least of order a microgauss. We have also estimated the proton flux at earth from the Virgo cluster. Assuming a few (1-10) % of the ram pressure of the infalling matter would be transferred to the cosmic-rays, the estimated flux for $E \sim 10^{19}$eV is consistent with observations, so that such clusters could be plausible sources of the UHE CRs.

*Subject headings:* Cosmic Rays - Hydrodynamics - Particle Acceleration

## 1. Introduction

One of the most puzzling problems in astrophysics is the origin of the ultra high energy cosmic rays (UHE CRs) of $E > 1 \text{EeV} (= 10^{18}$ eV) (Hillas 1984, Elbert & Sommers 1995). According to Wolfendale (1993) and Bird *et al.* (1994), *at least* some of these UHE CRs are likely to be protons, so an origin within our own galaxy runs into several serious problems. Also the slope of the observed particle flux flattens somewhat around $E = 3$ EeV (Teshima 1993) indicating that the origin of the UHE CRs might be different from that of lower energy CRs; thus signalling a possible transition from a galactic component to an extragalactic component. Models based on sources of the UHE CRs in the Local Supercluster are discussed by Berezinsky & Grigor'eva (1979), and Giler *et al.* (1980).

Berezinsky & Grigor'eva (1988, BG88 hereafter) showed that a bump in the UHE CR spectrum could be produced by energy losses due to the interaction of the UHE protons with the cosmic microwave background radiation (CMBR), if the UHE protons come from extragalactic sources; the Greisen effect (Greisen 1966; Zatsepin & Kuzmin 1966). Rachen & Biermann (1993) and Rachen *et al.* (1993) suggested that the UHE protons below the Greisen cutoff are accelerated diffusively at strong shocks in the hot spots of powerful radio jets. They estimated the energy available for UHE protons from the synchrotron luminosity of these radio sources, assuming the usual minimum energy condition. They were not, however, clear about how the protons accelerated at the hot spots escape from the jets. The escape process is likely to be energy dependent, especially if it is diffusive, so it will affect the spectrum of particles detected at earth.

Since it is now widely believed that CRs are diffusively accelerated at many shocks by the first-order Fermi process (see reviews by Drury 1983, Jones & Ellison 1991), it is natural to examine the strongest and biggest shocks in the astrophysical environment as possible acceleration sites for the UHE CRs. Those would be accretion shocks around galaxy clusters. The gravitational theory of the formation of large scale cosmological structure predicts that primordial density perturbations grow into pancake-like structures and that clusters form at the vertices where the pancakes intersect. Recent cosmological hydrodynamic simulations (e.g. Kang *et al.* 1994, Bryan *et al.* 1994, Cen & Ostriker 1994) have shown that some cosmological models produce structure consistent with X-ray observations of galaxy clusters. These simulations have shown that accretion shocks form around the caustic surfaces and that clusters of galaxies are surrounded by these accretion shocks. X-ray clusters bright enough to be observed (e.g. $L_x > 10^{43}$ ergs s$^{-1}$) typically have a mean temperature in the range 2-10 keV (Edge & Stewart 1991). Assuming that this is the postshock temperature ($T_{ps} = 1.3 \times 10^7$ K$[v_s/1000$ km s$^{-1}]^2$), the estimated shock velocity ranges between $(1.3 - 3) \times 10^3$ km s$^{-1}$. In the present paper, we explore the possibility that protons can be accelerated to high energies at these accretion shocks via the first-order Fermi process.

## 2. Basic Physics

Two key requirements of our model are the existence of accretion shocks around clusters of galaxies with characteristic incident gas speeds, $v_S \sim (1-3) \times 10^3$ km s$^{-1}$, and the existence of a turbulent magnetic field, $B \sim 1$ $\mu$G. In this section we will discuss them in detail.

## 2.1. Accretion Shocks around Galaxy Clusters

Fig. 1 shows a slice cut through a representative cluster found in the numerical simulation previously reported by Kang *et al.* (1994) (henceforth KCOR). The calculation was performed with the cosmological hydrodynamic code described in Ryu *et al.* (1993). The details of the model parameters and the numerical methods can be found in KCOR. However such details are not essential to this discussion, since accretion shocks around clusters are generic structures also found theoretically (Bertschinger 1985) and numerically, independent of the model cosmology details. But, to be specific wherever relevant, we use $H = 50$ km s$^{-1}$/Mpc, as in KCOR. The location of the shock shows up nicely in the temperature contour plot in Fig. 1 through a steep gradient around the central hot gas. The velocity map also shows the shock and the accretion of the surrounding matter onto the cluster. The line-cut along $y = 20$ Mpc also shows a typical flow pattern of a caustic structure. The estimated shock velocity is about 2400 km s$^{-1}$ for this cluster. The gas density in the core region is affected by the numerical resolution and could be much higher. The total bremsstrahlung luminosity of the simulated cluster shown in Fig. 1 is estimated to be $L_x \sim 2 - 3 \times 10^{44}$ ergs s$^{-1}$

As shown in Fig. 1 the temperature of the cluster gas stays nearly constant, while the density concentrates toward the center. Hence, the observed X-ray cluster represents only the central part ($r_c \sim 0.5 - 1$ Mpc for the cluster in Fig. 1) of the entire cluster, while the simulated structure suggests that the temperature remains constant further out, where the accretion shock is ($r_S \sim 5$ Mpc for the cluster in Fig. 1). The gas density and the velocity of infalling matter are dependent on the power and the wavelength of initial caustics that later form the clusters. So, the upstream density and the shock speed should be related to the size of the cluster. For X-ray bright clusters relevant to our discussion, the gas density upstream of the shock is typically $(0.1-0.01)\rho_{\rm crit}$ and the infalling velocity is $1000-3000$ km s$^{-1}$ in an $\Omega = 1$ CDM universe with baryonic density $\Omega_b = 0.06$ (KCOR). Here $\rho_{\rm crit}$ is the critical density of the universe ($4.7 \times 10^{-30}$ g cm$^{-3}$). The X-ray flux from the low density gas near accretion shocks cannot be detected, so direct observational proof of the existence of the accretion shock would be extremely difficult to obtain, if not impossible.

## 2.2. Magnetic Field and Diffusion Coefficient

The current observational upper limit for a large-scale, pervading intergalactic magnetic field is about $10^{-9}$G (Kronberg 1994). On the other hand, there is some speculation that magnetic fields in rough equipartition with the cosmic background radiation ($B \sim 1\mu$G) exist in intergalactic space in the local universe and that such fields may be turbulent (e.g., Kronberg 1994). According to many recent observations (e.g., Kim *et al.* 1991, Taylor *et al.* 1994, Feretti *et al.* 1995) magnetic fields near the $\mu$G level are common in cluster core regions ($r < 1$Mpc). Those observations indicated that the field strength decreases outward, however, so the field at the outskirts of clusters might not be as strong as in the core regions. On the other hand, Vallée (1990) detected a magnetic field component of $\sim 1.5 \mu$G in a 10 Mpc region around the Virgo cluster. Hence it is very plausible that there exist $\mu$G magnetic fields around the clusters of galaxies on a scale much broader than 1 Mpc, even though the large-scale field far away from the clusters is less than $10^{-9}$G. Since they are a necessary ingredient to produce UHE CR there, we will assume the existence of $\sim \mu$G magnetic fields near accretion shocks around galaxy clusters.

Diffusive, first-order Fermi acceleration at shocks depends on the existence of magnetic irregularities (usually ascribed to Alfvénic turbulence) capable of strongly scattering energetic protons. This keeps some superthermal protons in the vicinity of the shock for a long time and causes them to cross the shock discontinuity repeatedly, gradually gaining energy from the different speeds of the scattering centers across the shock. Theoretically the process is described by a spatial diffusion coefficient, $\kappa$. To estimate the diffusion coefficient we have to know not only the overall field strength but also the power spectrum of the field irregularities or the Alfvén waves. The scale of irregularities most important to scattering protons of a given energy is the gyroradius of those protons; i.e., $1/k \approx r_g \approx 1$ kpc $(E_{\rm E}/B\mu)$, where $B\mu$ is the magnetic field in units of $\mu$G and $E_{\rm E}$ is the proton energy in EeV. It is commonly assumed in calculations of diffusive shock acceleration that the scattering Alfvén waves are self-generated by the CRs themselves by the so-called "streaming instability" (Wentzel 1974). That may apply in this context only at energies lower than those of primary interest to us here, however. A rough estimate for the growth time scale for the waves (e.g., Skilling 1975) is $\tau_w \sim 10^{10}$ years $(E_{\rm E}/B\mu)$, longer than the age of the universe for $E > 1$ EeV even with $B\mu \sim 1$. On the other hand, the irregularities in the field within the cluster can be generated by various dynamical effects such as the supersonic motion of galaxies through ICM, galactic winds and galaxy mergers. The scattering lengths of interest are of order the size of individual galaxies, so such dynamical effects may very well produce substantial turbulence that is directly relevant. Some observations (Feretti *et al.* 1995) indicate that the ICM field is likely to be tangled on scales of the order of less than 1 kpc. Also, recent high-resolution numerical simulations indicate that the gas flows inside clusters become turbulent during the cluster formation (Cen 1995). Then turbulent flows may subsequently generate the irregularities in magnetic field. Outside the accretion shock we must hypothesize the existence of pre-existing field turbulence. Adiabatic compression of that field and turbulence by convergence of the inflow onto the accretion shock might very reasonably lead to an acceptable level of magnetic irregularities in the shock vicinity.

One can express the diffusion coefficient as $\kappa = \kappa_{\rm B} (B^2/\delta B^2)$, where

$$\kappa_{\rm B} = \frac{3.13 \times 10^{22} {\rm cm}^2 {\rm s}^{-1}}{B\mu} \frac{p^2}{\sqrt{p^2+1}}, \qquad (1)$$

is known as the Bohm diffusion coefficient and $(B^2/\delta B^2)$ represents the ratio of total magnetic energy to the energy in magnetic irregularities of the appropriate scale (e.g., Drury 1983). In this discussion particle momentum is expressed in units of $mpc$. Since the spectrum of field irregularities is unknown, we will for now take the simplest approach and assume Bohm diffusion; i.e., $(B^2/\delta B^2) \sim 1$.

### 2.3. Maximum Energy

The mean acceleration time scale for a particle to reach momentum $p$ is determined by the velocity jump at the shock and the diffusion coefficient (e.g., Drury 1983), that is,

$$\tau_{acc} = \frac{p}{<\frac{dp}{dt}>} == \frac{3}{u_1 - u_2}(\frac{\kappa_1}{u1} + \frac{\kappa_2}{u_2}) = \frac{8}{u_s^2}\kappa_B . \tag{2}$$

Here the subscripts, 1 and 2, designate the upstream and downstream conditions, respectively. The strong shock limit is taken, so $u_2 = u_1/4$, and we assume for a turbulent field that $B/\rho$ is constant across the shock, so $\kappa/u$ = constant. Using the Bohm diffusion coefficient in equation (1), the mean acceleration time scale is given by

$$\tau_{acc} = (3.5 \times 10^9 \text{years})(\frac{p}{10^{10}})B_\mu^{-1}(\frac{u_s}{1500 \text{km s}^{-1}})^{-2}. \tag{3}$$

This tells us that the diffusive acceleration time scale increases directly with the energy and that it takes a few billion years to accelerate the proton to $E = 10$ EeV at a typical cluster shock.

On the other hand, an UHE proton loses energy due to its interactions with the CMBR; that is, pair production and photopion production. This occurs continuously while it is being accelerated at the source, as well as during its travel through extragalactic space from the source to the earth. The energy loss rates of these processes are presented in detail in BG88. According to Fig. 1 in BG88 pair production is the dominant loss mechanism around and below 10 EeV. The energy loss timescale at 10 EeV due to this process is $\tau_{pp} \sim 5 \times 10^9$ years. Above the Greisen cutoff energy, $E \sim 60$ EeV, the photopion interaction becomes dominant and the loss timescale decreases rapidly and then approaches $\tau_\pi \sim 5 \times 10^7$ years for $E > 100$ EeV. This is much shorter than $\tau_{acc}$ for the energy range under consideration, so the photopion reaction limits the proton energy accelerated by the cluster shocks to energies below $E = 60$ Eev. Also, these time scales are for the present epoch. They decrease as $\tau(E, z) = (1 + z)^{-3}\tau([1 + z]E, 0)$, due to the increase of the photon number density and photon energy at high redshifts (BG88).

Considering that the cluster accretion shocks probably form later than $z < 5$ and that the photopion reaction inhibits the protons from achieving ultrahigh energies at large redshifts, we will focus on the recent epoch ($z < 1$) and ignore the cosmological evolution for the time being. Then the available period for acceleration is estimated to be $t_d \sim 0.6 t_{age}$, where the age of universe is $t_{age} = 1 - 2 \times 10^{10}$ years. During this time, for $E < 60$ EeV, the maximum proton energy is set by the condition that $\tau_{acc} = \tau_{pp}$, that is, the acceleration time scale equals the loss time due to pair production reaction. Then the maximum momentum can be written as

$$p_{\max} \simeq \frac{E_{\max}}{m_p c^2} \simeq 1.4 \times 10^{10} \left(\frac{u_s}{1500 \text{km s}^{-1}}\right)^2 B_\mu. \qquad (4)$$

For simplicity, here we assume a constant shock speed. According to the analytic study of self-similar cluster evolution (Bertschinger 1985), the velocity scales as $t^{-1/9}$, so a constant shock speed should be a good approximation. So for $u_s \sim 3000$ km s$^{-1}$ and $B_\mu \sim 1$, the maximum energy $E_{\max} \sim 56$ EeV, equaling the Greisen cutoff energy can be reached by protons through diffusive acceleration at cluster shocks. We note here that heavy nuclei can also be accelerated at the cluster shocks, and that for a given momentum, Bohm diffusion leads to more rapid acceleration for larger nuclear charges. So those nuclei might possibly reach higher energies than the Greisen cutoff.

## 3. Virgo Cluster Model

The Virgo cluster, $\sim 17$ Mpc away from us, is the nearest X-ray bright cluster (Freeman *et al.* 1994). So, to provide an indication of how significant the flux from such clusters might be if they are able to generate particles of sufficient energy, we estimate the observed flux of UHE CRs from the Virgo cluster shock. We make simple assumptions about the physical parameters of the cluster shock, and the magnetic field near the shock and in intergalactic space. In addition, as Rachen & Biermann did, we assume UHE CRs propagate rectilinearly once they escape into the intercluster space. However, we do include a treatment of the escape itself. This discussion is meant to be an illustrative exercise rather than a rigorous derivation.

The core radius of Virgo is 0.2 Mpc and the 2 − 10 keV luminosity is about $2 \times 10^{43}$ergs s$^{-1}$, so it is relatively small compared to rich clusters. The optical image is $\sim 12°$ in diameter and the detectable X-ray emitting region is as broad as the optical extent of the cluster (Takano *et al.* 1989). So the radius of the detectable X-ray cloud is $\sim 1.8$ Mpc. The mean temperature of the cluster gas is 2.4 keV (Edge & Stewart 1991) which in terms of a postshock temperature translates into $v_s = 1.5 \times 10^3$ km s$^{-1}$. The magnetic field in a 10 Mpc region around the Virgo cluster is about 1.5 $\mu$G according to observations by Vallée (1990).

Using the results from KCOR we assume that a spherical shock of radius 5 Mpc surrounds the cluster and that the speed of the flow coming into the shock is 1500 km s$^{-1}$. The proton momentum distribution at the shock is assumed to be a power law whose index is 4, as expected for a strong shock, and with an exponential cutoff, so

$$f_{sh}(p) = f_o\, p^{-4} \exp\left[-\left(\frac{p}{p_{\max}}\right)\right]. \qquad (5)$$

The minimum energy for the power law is $p_{\text{inj}} \sim (u_s/c)$ and the cutoff energy estimated by equation (4) is $E_{\max} = 14$EeV.

The normalization constant $f_O$ is determined by assuming that a small fraction of the ram pressure ($\rho_1 u_s^2$) is converted to CR energy (see, e.g., Kang & Jones 1990). Here the infall gas density is $\rho_1 = (0.01 - 0.1)\rho_{\rm crit}$. Instead of having two parameters ($\rho_1/\rho_{\rm crit}$ and $E_{cr}/(\rho_1 u_s^2)$), we put them together into a single parameter to normalize the postshock CR energy density as a fraction of $\rho_{\rm crit} u_s^2$; that is, $\epsilon = E_{cr}/(\rho_{\rm crit} u_s^2)$. Thus $f_O$ is calculated approximately according to

$$E_{cr} \simeq 4\pi m_p c^2 \int_{p_{\rm inj}}^{p_{\rm max}} f_O\, p^{-4} \left(\sqrt{p^2+1} - 1\right) p^2 dp \simeq 4\pi m_p c^2 f_O \ln\left(\frac{p_{\rm max}}{p_{\rm inj}}\right) \qquad (6)$$

The particle distribution given by equation (5) represents the particles confined around the shock that continue to be accelerated. The particles observed at earth, however, are those that escape from the shock into extragalactic space and travel to us either diffusively or rectilinearly once they exit the cluster vicinity. Since the infalling matter converges onto the spherical shock, the particles escape to us from the upstream region. The particle distribution upstream ($r > r_s$) of the shock decreases exponentially on a diffusion length scale (Drury 1983) as

$$f(r,p) = f_{sh}(p) \exp\left[\frac{-(r-r_s)}{l_d(p)}\right] \qquad (7)$$

where the diffusion length for the momentum $p$ is $l_d(p) = \kappa(p)/u_s$. For Bohm diffusion $l_d \sim 0.9 \text{Mpc}(u_s/1500 \text{ km s}^{-1})^{-1}$ for $E_{\rm max}$. Since the low energy particles have smaller diffusion lengths, they are mostly confined at the shock. Only the high energy particles can diffuse away from the shock and escape in significant numbers. So the spectrum of escaping particles is dependent both on the diffusion coefficient and the escape process.

We are counting on a pre-existing magnetic turbulence outside the cluster shock. We can assume that it will be advected into the shock during accretion, and if it is strong that $(B^2/\delta B^2)$ will remain relatively unchanged. In that case $\kappa \propto B^{-1} \propto \rho \propto ur^2$. Since the inflow speed relative to the shock does not change very fast with upstream distance, once the Hubble flow is taken into account (Bertschinger 1985), this leads to an approximate proportionality, $l_d \propto r^2$. Hence, because of the already large diffusion length found above, the highest energy particles should escape the system once they diffuse upstream to a distance comparable to the radius of curvature of the shock, since their lateral diffusion will become comparable to their radial diffusion (e.g., Eichler 1981). We assume for simplicity that particles effectively escape from the shock at the radial distance from the cluster center $r = 2r_s$, so the escaping particle distribution is $f_{\rm esc}(p) = f_{sh}(p) \exp\left[-r_s/l_d(p)\right]$. The shell from which the particles escape has an angular radius about 35°, which is much larger than the optical extent of the Virgo cluster. Then we further assume that the escaped particles travel to us in a straight line without further deflection, since the magnetic field in extragalactic space away from clusters is estimated to be lower than $10^{-9}$ Gauss (Kronberg 1994). The flux of highly relativistic particles ($v \sim c$) observed at the earth is then

$$F(p) = f_{\rm esc}(p)\, \frac{c}{4} \left(\frac{2r_s}{d_{\rm Virgo}}\right)^2 \qquad (8)$$

where c is the speed of light and the distance to the Virgo cluster is $d_{\text{Virgo}} = 17$ Mpc.

Fig. 2 shows the estimated particle flux $J(E)E^{2.75}$ in units of GeV$^{1.75}$cm$^{-2}$s$^{-1}$str$^{-1}$, and $J(E)E^3$ in units of eV$^2$cm$^{-2}$s$^{-1}$str$^{-1}$. Here $J(E)dE = F(p)p^2 dp$. The model parameters are $B_\mu = 1$, and $\epsilon = 3 \times 10^{-3}$ (3% to 30% efficiency in conversion of inflowing energy to cosmic ray energy by the shock). The parameter $\epsilon$ determines the normalization. The overall shape of the spectrum observed from this individual cluster is regulated by the exponential factor controlling escape, $\exp[-r_s/l_d(p)]$, for the lower energy particles and by the exponential acceleration cutoff at $E_{\text{max}}$ at higher energies. One can see from Fig. 2 that near 30 EeV the particles from the Virgo cluster could contribute a substantial fraction of the observed flux, but the detailed shape of the spectrum depends on the assumed escape process. Note here that the observed flux below the "ankle" at 3 EeV could be dominated by a different (possibly galactic) component, and that contributions from other distant clusters of galaxies would make the "cluster shock" component broader than the curve shown in the figure, since those protons would loose energy in transit. In this calculation, we did not include energy losses for protons in flight to earth, since the Virgo cluster is not far way. According to Fig. 4 of BG88, proton energy losses are insignificant for energies less than 200 EeV for the propagation time of $\tau = 5 \times 10^7$ years ($\sim 17\text{Mpc}/c$).

We might also expect electrons to be accelerated in the cluster accretion shocks. The energy of those electrons will be much more severely limited, however, by synchrotron and Compton scattering losses (which will be comparable). Since both electrons and protons are highly relativistic, it is reasonable to expect them to propagate according to the same diffusion coefficient. By equating the acceleration time to the cooling time through synchrotron emission, one can show from standard formulae that the maximum electron Lorentz factor would be $\gamma \sim 10^8 u_s / B_\mu^{3/2}$, which would be achieved in $\sim 10^5 u_s^2 / B_\mu$ years. These highest energy electrons would radiate at soft X-ray energies from a thin shell around the shock. They would cool much too quickly to fill the cluster. For that reason the X-ray power fails by many orders of magnitude being detectable, however. Electrons of much lower energy, say $\gamma \sim 10^4$, would radiate at radio wavelengths and could conceivably fill the cluster, since their lifetimes are $\sim 10^9$ years. If they uniformly fill the cluster at this energy the integrated flux at a few hundred MHz could be $\sim 1$ Jy, if the efficiency of electron acceleration is similar to that assumed for protons; i.e., $\epsilon_e \sim 10^{-3}$. This might seem detectable at first glance, but the flux would be spread over $\sim 0.3$ sterad, so the surface brightness is far too small to distinguish from various backgrounds.

## 4. Conclusion

In this paper, we have argued that UHE protons may be accelerated up to the energies just below the Greisen cutoff at accretion shocks expected around clusters of galaxies by the first-order Fermi process, if turbulent magnetic fields $\sim \mu$G exist around the shocks. Proton energies can be increased only to the Greisen cutoff near $E = 60$ EeV by the proposed mechanism, since above that energy losses due to photopion reaction with the CMBR are much faster than the energy gain

– 9 –

due to diffusive acceleration. According to the hydrodynamic simulation by KCOR, the accretion shocks are about 5 Mpc from the cluster center and, thus, in a larger and more tenuous region than that made apparent by the galaxy distribution or X-ray emission from hot intracluster medium gas. Those are typically within 0.5 − 1 Mpc of the cluster center. Existing observations indicate not only that the magnetic field strength in the core of some Abell clusters is indeed about a $\mu$G (Kim *et al.* 1991), but also that a field of similar strength may extend much further out (5 Mpc in radius) in the Virgo cluster (Vallée 1990).

To provide a concrete example we have estimated the observed UHE CR flux expected from the shock around the Virgo cluster, the nearest X-ray bright cluster. We assumed the cluster shock speed is 1500 km s$^{-1}$ and a magnetic field of $\sim \mu$ G exists around the shock, and (3-30)% of the kinetic energy of the infalling matter is converted to CR at the shock. Assuming a simple escape process in which the particles escape rectilinearly from the cluster once they diffuse significantly beyond the shock, we found that a substantial fraction of the observed UHE CR flux at the energies near 30 EeV could come from the Virgo cluster. On the other hand, we expect contributions from other, more distant X-ray clusters could give the particle flux which is consistent with the observed flux on a broader energy range (but still below the Greisen cutoff). In the future we will examine the ways that such sources would contribute to the observed UHE CR flux, including the integrated particle spectrum from them and the issue of relatively isotropic arrival directions. Rachen & Biermann (1993) in their model for UHE CR, based on the most powerful radio jets also found that a few of the closest potential sources might individually be significant contributors. Thus, it may be that all discrete source models like these would lead us to expect significant granularity to eventually be observationally apparent.

In order to understand properly what role cluster shocks might play in the production of high energy cosmic rays and associated electromagnetic radiation further detailed studies are needed to clarify issues concerning magnetic field structure, particle diffusion and escape as well as propagation of the particles through extragalactic space.

We thank an anonymous referee for helping us recognize important issues regarding the photopion energy loss and intergalactic magnetic field. We thank D.-W. Kim for helpful discussion on X-ray cluster observations. M. Teshima generously provided us with the observed CR energy spectrum. The work by HK was supported in part by the Korea Research Foundation through the Brain Pool Program. The work by DR was supported in part by the Non-Directed Research Fund of the Korea Research Foundation 1993. The work by TWJ was supported in part by the NSF (AST-9318959), by NASA (NAGW-2548) and by the Minnesota Supercomputer Institute.

Fig. 1.— Gas temperature contour plot (top left) and velocity map (top right) of a slice with 0.6 Mpc thickness for a representative X-ray bright cluster in a simulation of large scale structure. Bottom three plots are for the gas density, temperature and velocity in $x$-direction for a line cut at $y = 20$ Mpc along $x$-axis of the same cluster.

Fig. 2.— Estimated proton flux from the Virgo cluster assuming that the accretion shock speed is 1500 km s$^{-1}$, the magnetic field strength is 1 $\mu$G, and $3 \times 10^{-3} \rho_{\rm crit} u_S^2$ is transferred into CR energy density. The observed points (cross: Fly's Eye, filled circles: Akeno, open circles: Havera Park, open squares: Yakutsk) in $J(E)E^3$ plot are taken from Teshima (1993).

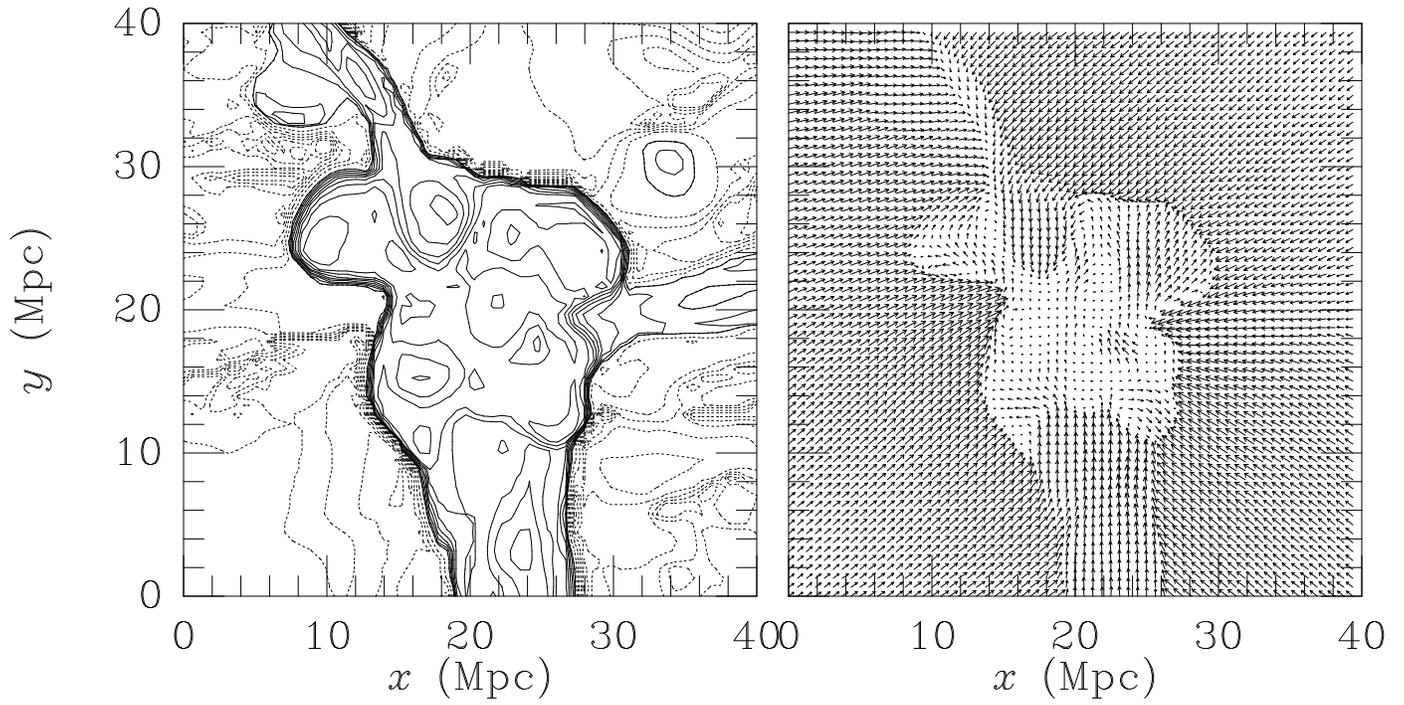
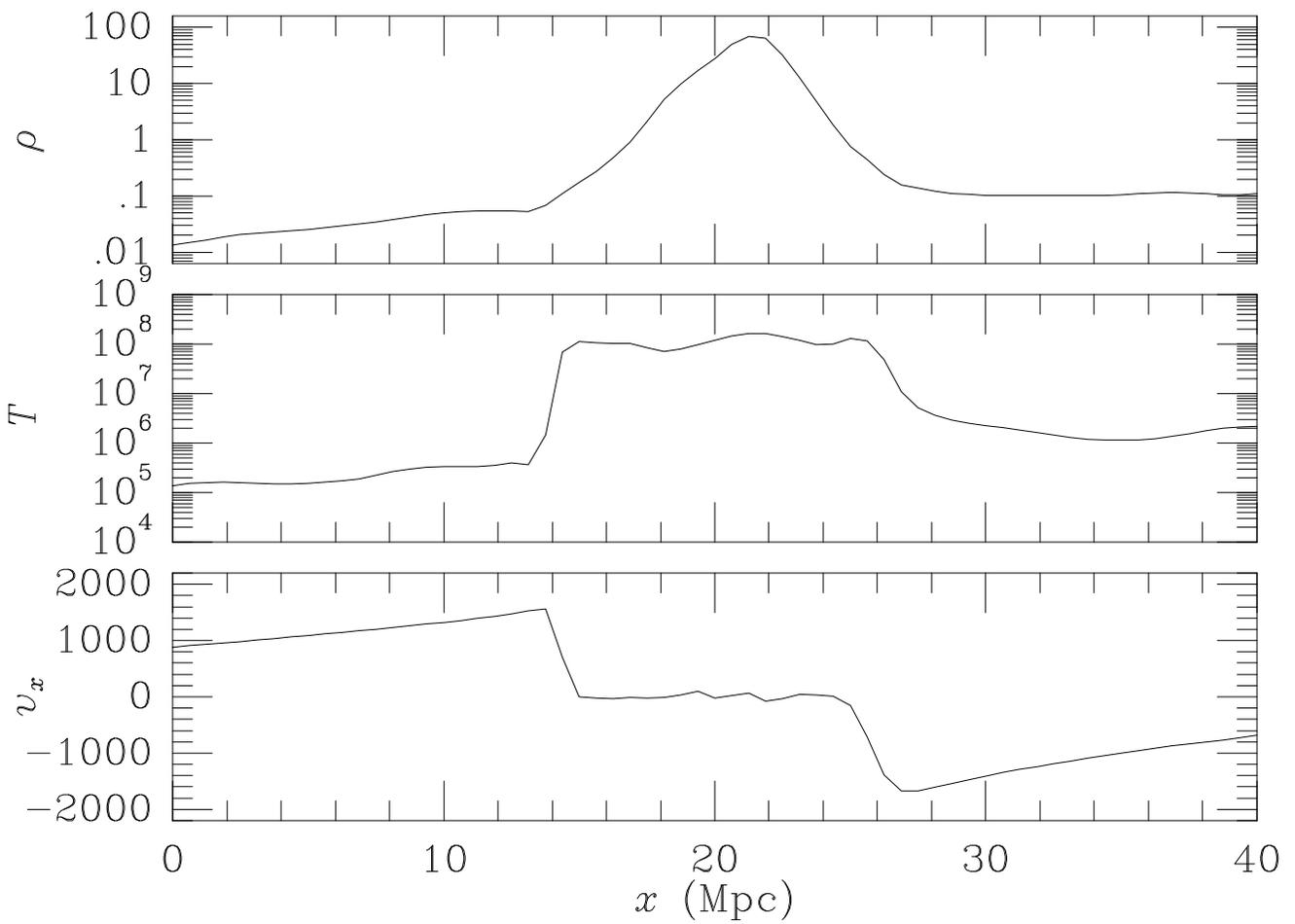

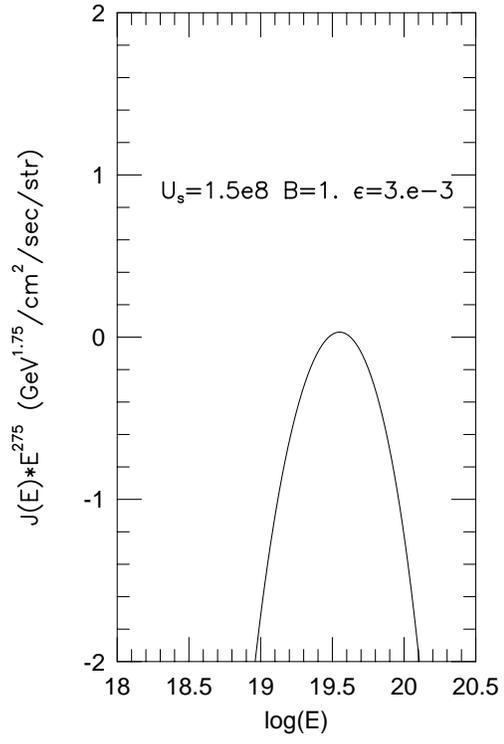 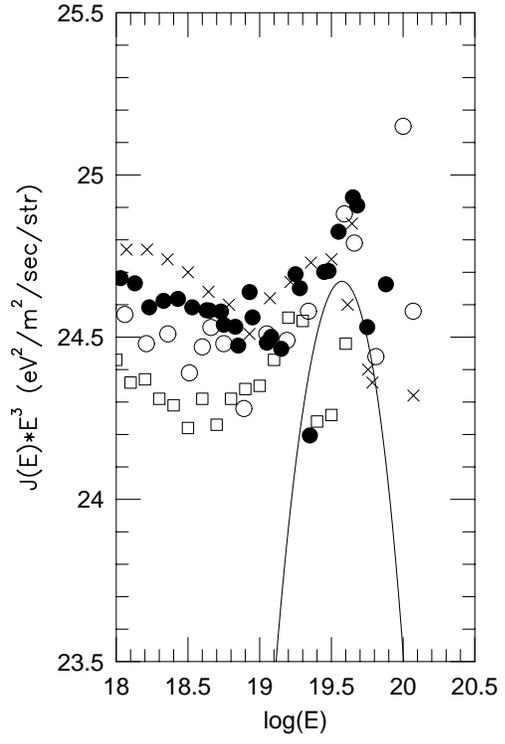